\documentclass[fleqn,10pt]{wlscirep}
\usepackage[colorinlistoftodos]{todonotes}
\usepackage{chemformula}
\usepackage{amsmath}
\usepackage{indentfirst}
\usepackage{subfig}
\usepackage{graphicx}
\usepackage{braket}
\usepackage{soul}
\DeclareUnicodeCharacter{2212}{-}
\usepackage[section]{placeins}
\let\sub\textsubscript

\title{Analytical Models of Phonon-Point Defect Scattering}

\author[1]{Ramya Gurunathan}
\author[1]{Riley Hanus}
\author[1]{Maxwell Dylla}
\author[2]{Ankita Katre}
\author[1]{G. Jeffrey Snyder}

\affil[1]{Materials Science and Engineering, Northwestern University, Evanston, IL, 60208, USA}

\affil[2]{Centre for Modelling and Simulation, Savitribai Phule Pune University, Pune, Maharastra, India}
\affil[*]{jeff.snyder@northwestern.edu}

\begin{abstract}
Point defects exist widely in engineering materials and are known to scatter vibrational modes to reduce thermal conductivity. The Klemens description of point defect scattering is the most prolific analytical model for this effect. This work reviews the essential physics of the model and compares its predictions to first principles results for isotope and alloy scattering, demonstrating the model as a useful materials design metric. A treatment of the scattering parameter ($\Gamma$) for a multiatomic lattice is recommended and compared to other treatments presented in literature, which have been at times misused to yield incomplete conclusions about the system's scattering mechanisms. Additionally, we demonstrate a reduced sensitivity of the model to the full phonon dispersion and discuss its origin. Finally, a simplified treatment of scattering in alloy systems with vacancies and interstitial defects is demonstrated to suitably describe the potent scattering strength of these off-stoichiometric defects.

\end{abstract}

\begin{document}

\flushbottom
\maketitle
%
%
\thispagestyle{empty}

\section{Introduction}

Modelling the lattice thermal conductivity, or the heat transported through atomic vibrations, has long been important to a wide range of science and engineering applications including thermoelectrics, thermal barrier coatings, and thermal management in electronic materials. All of these functional materials rely on doping and alloying to tune their properties, and so the impact of impurities and other point defects on the lattice thermal conductivity is important to understand\cite{Wang2012, Toberer2011}.   

Peierls presented one of the earliest solutions for lattice thermal conductivity in 1929 by evaluating the phonon Boltzmann transport equation, which was simplified by Callaway based on the relaxation time approximation. Later, Klemens established a theory for vibrational mode scattering due to static imperfections in a lattice, and provided closed-form expressions for thermal conductivity versus defect concentration still utilized today\cite{Klemens1955, Klemens1960, Goldsmid2007}. These analytical expressions based on low order perturbation theory are attractive because of their simplicity and utility for determining the source of phonon scattering and thermal conductivity suppression in a system. By calculating the relative contribution of independent scattering mechanisms such as mass disorder and local strain effects, one can determine the dominant mechanisms of scattering in a defective system and uncover material design strategies to optimize the thermal conductivity for a given technological application\cite{Meisner1998, Yang2015, Wood2018, Toberer2010}. 

First-principles techniques have been developed recently to compute the impact of point defects on thermal transport. These simulations have shown very good quantitative agreement with experiments for a range of materials and have provided useful insights regarding the mechanisms of phonon-defect scattering\cite{Katre2016, Katre2018}. However, multiple calculations are required to compute defect structures, evaluate scattering strengths, and solve the Boltzmann transport equation for the thermal conductivity\cite{Li2012, Katre2017, Polanco2018, Mingo2010}. Often, these techniques are too expensive and system dependent for routine modelling used to determine the dominant scattering mechanisms in a system\cite{Schrade2018}. While the first-principles methods are essential to understanding vibrational mode properties, and in many cases elucidate limitations of analytical phonon theories, the Klemens point defect model has proven to be highly descriptive across material systems and therefore continues to be widely used\cite{Polanco2018, Schrade2018, Shiga2014, Feng2015}.

The Klemens equations are defined within the ostensibly limiting approximation of a single atom unit cell and the Debye model, or linear phonon dispersion. However, by comparing to both first-principles results as well as experiment, the predictive quality of this model is demonstrated even for complex unit cell materials.  

This paper provides a functional guide for understanding the influence of point defects on phonon transport and applying the Klemens equations to model thermal conductivity data. In addition, it resolves discrepancies between popular representations of the mass difference model, which have led to consistent errors in model inputs that may yield large factor differences in the predicted lattice thermal conductivity ($\kappa\sub{L}$). This study also re-evaluates the limitation of these equations to the Debye model dispersion. A mechanism is demonstrated for how the model's sensitivity to dispersion relation is, in practice, lifted, justifying the use of the model in systems with arbitrary dispersion relations.

The Klemens model predicts the ratio of the defective solid's lattice thermal conductivity to that of a reference pure solid ($\kappa\sub{L}/\kappa_0$). The basic functional form of the ratio is $\mathrm{tan^{-1}}u/u$, where the disorder scaling parameter $u$ is related to the pure lattice thermal conductivity reference ($\kappa_0$), elastic properties of the host lattice through its speed of sound ($v\sub{s}$), the volume per atom ($V_0$), as well as a scattering parameter ($\Gamma$), which captures the lattice energy perturbation at the defective site

\begin{align}
\frac{\kappa\sub{L}}{\kappa_0} = \frac{\mathrm{tan}^{-1}u}{u}
\hspace{20 mm}  
u^2 = \frac{(6\pi^5V_0^2)^{1/3}}{2 k_B v_s}\kappa_0 \Gamma.
\label{eqn:kappa_atan}
\end{align}

\begin{figure*}
\centering
\includegraphics[width=0.6\textwidth]{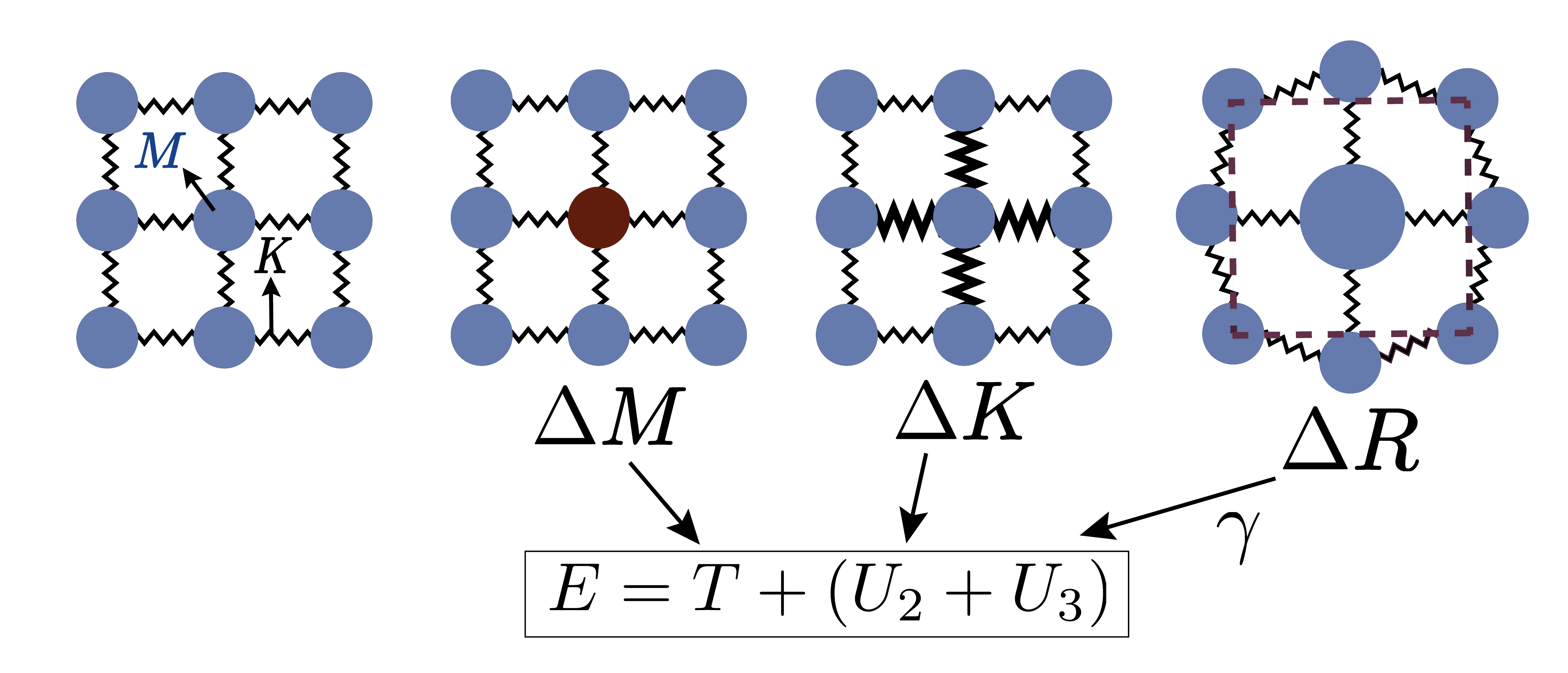}
\caption{The lattice perturbation mechanisms of a point defect include a mass difference ($\Delta M$), harmonic force constant difference ($\Delta K$), and strain scattering from site radius difference ($\Delta R$). Each contribution perturbs the lattice Hamiltonian ($E$) through a different term. $T$ is the kinetic energy of the lattice, and $U_{\mathrm{2}}$ and $U_{\mathrm{3}}$ are the harmonic and anharmonic contributions to the lattice potential energy.}
\label{fig:PDmech}
\end{figure*}

At each composition, the values of $\kappa_0$, $v\sub{s}$, and $V_0$ are linearly interpolated between the end-member properties. The perturbation caused by point defects in a lattice can be understood as a combination of a kinetic energy perturbation due to the mass difference on the defect site ($\Delta M$) and a potential energy distortion due to both the harmonic force constant difference ($\Delta K)$ and a the structural distortion of nearest neighbors around the defect introduced by a site radius difference ($\Delta R)$ (see Figure \ref{fig:PDmech}). It is often the case that mass difference is the dominant effect, since large volume differences are energetically unfavorable for substitutional defects. For simplicity, the remaining equations in this section will be defined for mass difference scattering alone ($\Gamma\sub{M}$), but analogous expressions for the potential energy terms are discussed in later sections. 

The $\Gamma\sub{M}$ parameter is the average mass variance in the system ($\langle \overline{\Delta M^2} \rangle$) normalized by the squared average atomic mass ($\langle \overline{M} \rangle^2$)\cite{Berman1956, Slack1962, Yang2015, Tamura1984}. Note that in the notation below, site averages are denoted by a bar while stoichiometric averages are denoted by angular brackets ($\langle \rangle$).

\begin{align}
     \Gamma\sub{M} = \frac{\langle \overline{\Delta M^2} \rangle}{\langle \overline{M}\rangle^2}.
       \label{mass_variance}
\end{align}

In a compound, these averaged quantities are most easily calculated by treating each component of the compound separately. For example, a generic compound can be expressed with the formula unit: $A1_{c_1}A2_{c_2}A3_{c_3}...An_{c_n}$ (e.g. \ch{CaZn_2Sb_2}), where A$n$ refers to the $n$\textsuperscript{th} component (e.g. Ca, Zn, or Sb) and c$_{n}$ refers to the stoichiometry of that component (e.g. 1, 2, or 2). 

For each site $n$ in the compound, Equation \ref{eqn:M2n} gives the average mass variance ($\overline{\Delta M^2}_n$) and average atomic mass ($\overline{\Delta M^2_n}$) specifically for that site, which can be occupied by a set of atomic species $i$, including the host atom and any substitutional defects. 

\begin{align}
    \overline{\Delta M^2_n} = \sum_i f_{i,n} (M_{i,n} - \overline{M_n})^2
    \hspace{20mm}
    \overline{M_n} = \sum_i f_{i,n} M_{i,n}
    \label{eqn:M2n}
\end{align}

$\overline{\Delta M^2_n}$ is defined by a sum over $i$ of the species site fraction ($f_{i,n}$) multiplied by the mass variance at each defect site, defined from the species mass $M_{i,n}$ and average atomic mass at that site $\overline{M_n}$\cite{Berman1976}. In vacancy scattering, where the perturbation emerges from both missing mass ($M\sub{vac}$) and missing bonds to nearest neighbors, a virial theorem derivation (see Section \ref{sct:vacancy}) suggests that the mass difference at the vacancy site should be $M_{i,n} - \overline{M_n} = -M\sub{vac} - 2\langle \overline{M} \rangle$. Finally, to derive the mass difference scattering parameter $\Gamma\sub{M}$, the stoichiometric averages of the $\overline{\Delta M^2_n}$ and $\overline{M_n}$ values are taken (Equation \ref{Gamma})\cite{Berman1976}.

\begin{align}
     \langle \overline{\Delta M^2} \rangle = \frac{\sum_n c_n \Delta M^2_n}{\sum_n c_n}
     \hspace{20mm}
     \langle \overline{M} \rangle = \frac{\sum_n c_n \overline{M_n}}{\sum_n c_n}
     \label{Gamma}
\end{align}

\section{Comparison to the Popular Mass Difference Expressions}

The mass difference model captured in Equation \ref{eqn:M2n} and \ref{Gamma} is a reformulation of the popular equation proposed in Yang \textit{et al.}\cite{Yang2015}, and follows the interpretation of Berman, Foster, and Ziman\cite{Berman1956}. It is recommended for conceptual clarity. This section reviews other popular interpretations of the mass difference model to understand the conceptual differences and compare the numerical results.

\subsection{Tamura Model}
\label{sct:Tamura}

The mass difference model proposed by Tamura preserves the dependence of the phonon relaxation times on polarization vector and the spatial anisotropy of atomic sites within the primitive unit cell, and is frequently implemented in numerical Boltzmann transport equation solvers for thermal conductivity\cite{Tamura1983, Tamura1984, Shiga2014, Feng2015, Polanco2018, Larkin2017, Li2012, Togo2015, Carrete2019, Li2014}. The mass difference parameter in the Tamura model ($\Gamma^\mathrm{T}\sub{M}$) is given as a sum over all the $s$ atom sites in a simulation cell, where $i$ again labels the species that may occupy site $s$, including the host and impurity atoms. In a similar fashion to previous expressions, $M_{i,s}$ and $\overline{M_s}$ indicate the $i$\textsuperscript{th} species mass and the average mass on atomic site $s$, respectively. In this case, however, the mass difference term is weighted by the eigenvector components corresponding to atom $s$ in the incident ($\mathbf{e_k}(s)$) and final ($\mathbf{e_{k'}}(s)$) vibrational mode.  

\begin{equation}
\Gamma^\mathrm{T}\sub{M} = \sum_s \sum_i f_{i,s}(\frac{M_{i,s} - \overline{M_s}}{\overline{M_s}})^2|(\mathbf{e_k}(s)\cdot \mathbf{e_{k'}}(s))|^2
\label{eqn:Tamura}
\end{equation}

The eigenvectors are composed of the displacement vectors ($\mathbf{u}(\mathbf{k}, s)$) of each atomic site as it participates in a vibrational mode, weighted by the square root of the atomic mass ($\mathbf{e_k} = [\sqrt{M_1}\mathbf{u}(\mathbf{k}, 1) \ldots \sqrt{M_s}\mathbf{u}(\mathbf{k}, s)]$), and are finally normalized such that $|\mathbf{e_k}|^2 = 1$. These eigenvectors can be calculated from the DFT (Density Functional Theory) force constant matrix\cite{Deangelis2018}. The description of mass difference scattering here is general enough in its formalism that it could be used to describe the perturbation induced to a vibrational mode regardless of its spatial extent. Therefore, in addition to plane wave phonons, the vibrational modes of diffusons, locons, and propagons within the Allen and Feldman formalism are describable within the same point defect scattering theory\cite{Allen1999}. 

\begin{figure}
    \centering
    \includegraphics[width = \textwidth]{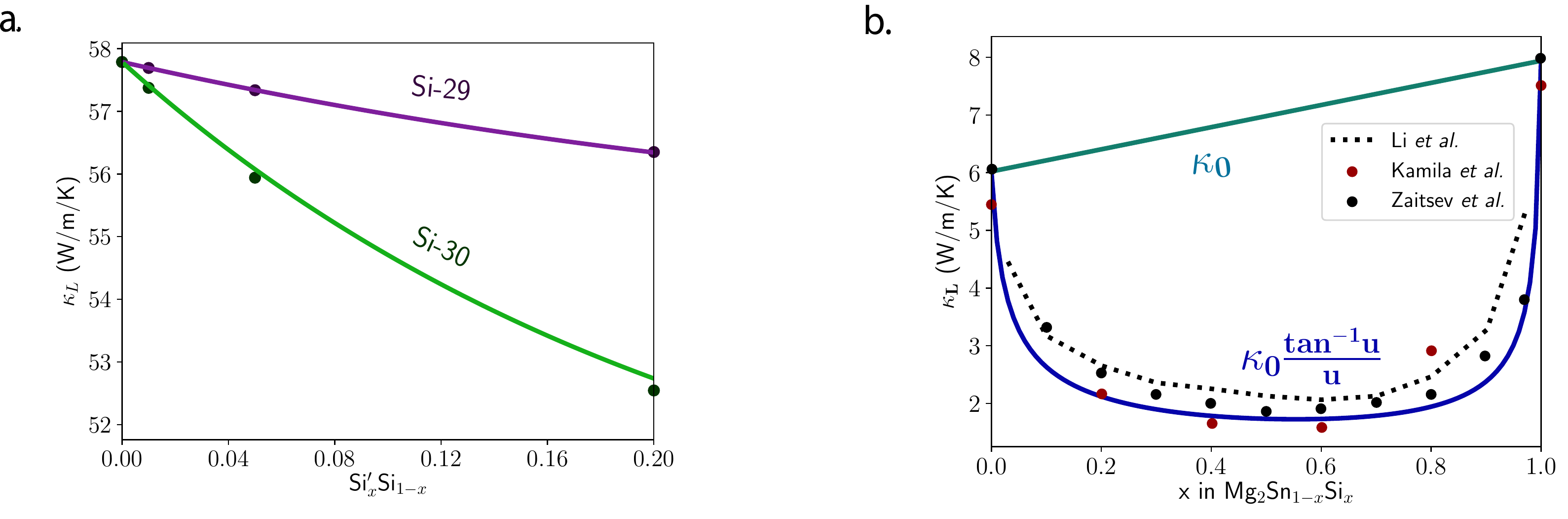}
    \caption{Point defect scattering from \textbf{T} matrix scattering theory (points) and the Klemens model (lines) for (a) Si isotope scattering (at 800 K) and (b) \ch{Mg_2Sn_{1-x}Si_x} (at 300 K)\cite{Li2012}}
    \label{fig:fp_comp}
\end{figure}

Point defect scattering has been studied with first principles techniques by applying DFT to compute the full vibrational spectrum, using T-matrix scattering theory and the Tamura model to compute point defect scattering rates, and finally solving the linearized Boltzmann transport equation to get $\kappa\sub{L}$\cite{Polanco2018, Li2012, Katre2017, Seyf2017, Katre2018, Katre2015}. In several reported materials systems, an excellent correspondence in shown between the results attained from first principles methods described above and the analytical Klemens model (Figure \ref{fig:fp_comp})\cite{Li2012, Tian2012}. It is important to remember that the Klemens model is fit to the end member thermal conductivity values, but still adequately predicts the suppression in thermal conductivity with compositional variation.

In addition to the Tamura model with full structural and lattice dynamical dependence, a closed form expression is presented for the low frequency limit, which depends only on the atomic masses. The assumption made here is that the displacement ($\mathbf{u}$) of each atom in a low-frequency mode is roughly equal in magnitude; therefore, one can assume the magnitude of an eigenvector element is simply proportional to the square root of the atomic mass ($|\mathbf{e}(\mathbf{k}, s)| \propto \sqrt{M_s}$). Following this assumption, Equation \ref{Gamma} can be derived, which is detailed in Supplemental Section \ref{supp:lf}.

\subsection{Primitive Unit Cell Model}
Finally, several texts, including original descriptions by Klemens, suggest defining all values of the mass difference model on the basis of the primitive unit cell\cite{Berman1956, Klemens1955, Klemens1957, Goldsmid2007, Goldsmid2010}. Physically, this treatment suggests a monatomic lattice approximation in which the atoms of the unit cell are simply summed together in a single large, vibrating mass. The scattering strength of the lattice can be thought of as an ensemble average of its microstates, or the primitive unit cells which compose it. Therefore, the unit cell model mass scattering parameter ($\Gamma^\mathrm{uc}\sub{M}$) can be calculated from the fraction ($P_c$) of unit cells with a mass of $M_c$ and their deviation from the average unit cell mass ($\overline{M}$). Finally, the mass differences are summed over all possible microstates in the lattice

\begin{equation}
 \Gamma^\mathrm{uc}\sub{M} = \sum_{c} P_{c} \left(\frac{M_c - \overline{M}}{\overline{M}}\right)^2 .  
\end{equation}

\begin{figure}
    \centering
    \includegraphics[width=0.6\textwidth]{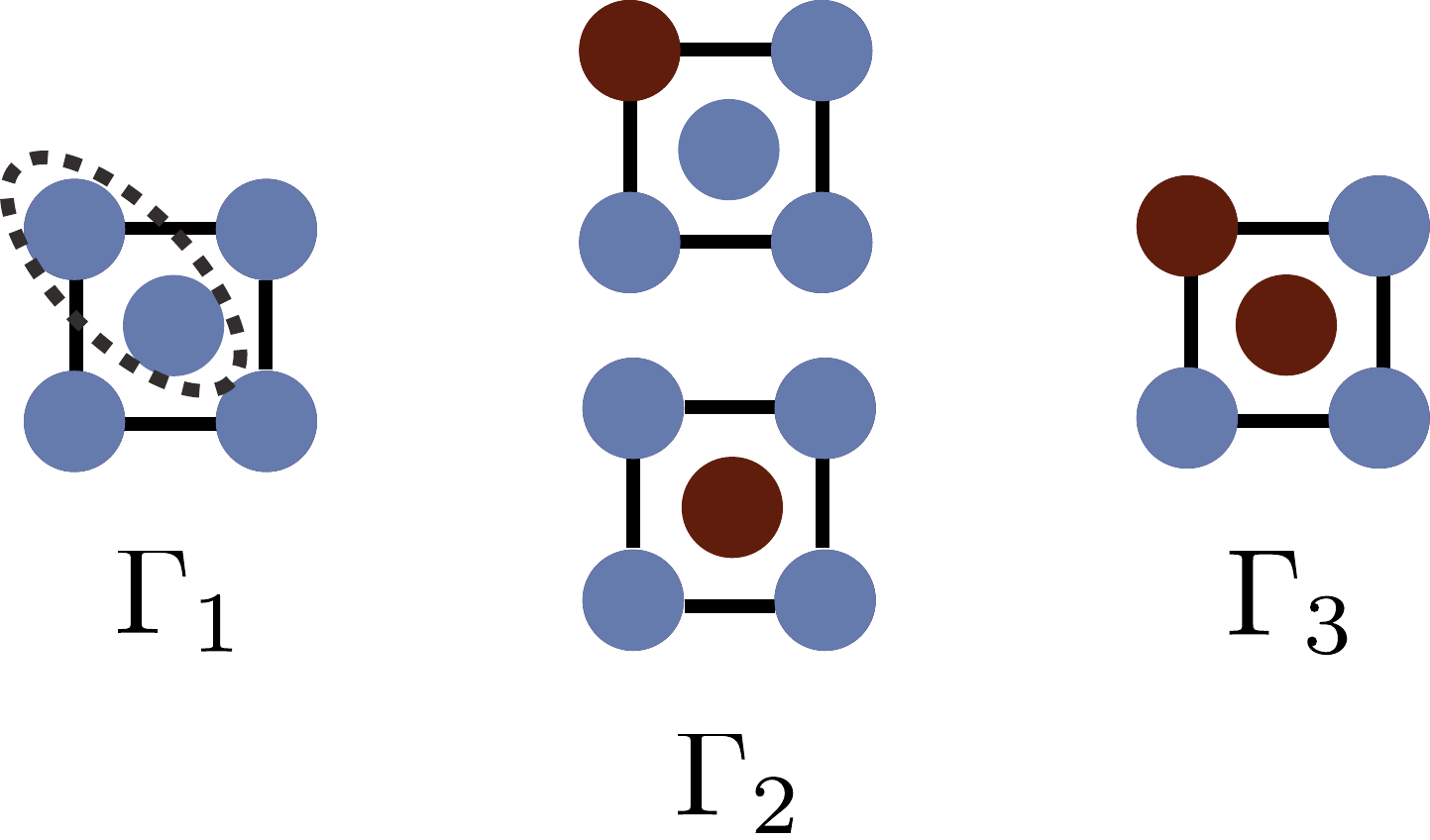}
    \caption{In an example 2-atom primitive unit cell (shown in dotted line), three possible microstates exist, containing 0,1, or 2 impurity atoms. In the unit cell basis, each microstate would contribute a term to the overall scattering parameter ($\Gamma$).}
    \label{fig:PUC}
\end{figure}

While most model inputs are well-defined, it is not immediately clear what the fraction of unit cells ($P_c$) should be. When there is a random distribution of defects in the lattice, the fraction or probability of finding a unit cell of mass $M_c$ can be determined from a binomial distribution function (described in more detail in Supplementary Section \ref{supp:puc}). A schematic of possible unit cells (microstates) for a 2-atom unit cell are shown in Figure \ref{fig:PUC}. In the case where all microstates are equally likely, the results are exactly equivalent to those produced by Equation \ref{Gamma} in the Introduction, which is more easily implemented. The benefit of the primitive unit cell interpretation arises when defect complexes are present in the lattice. Recent studies have identified these defect complexes in important engineering materials including clustering of Na dopant atoms in PbTe, antisite defects occurring in close proximity in BAs, vacancy clusters in crystalline Si, or Schottky and Frenkel defect pairs in functional oxides \cite{Wang2018,Lee2011,Zheng2018}. In these scenarios, the microstates corresponding to the defect complexes can be preferentially weighted with a larger fraction ($P_c$).

\subsection{Inconsistent Usage}

The dual formalisms of the scattering parameter $\Gamma$ based on either ``per unit cell" or ``per atom" quantities have led to inconsistencies in the calculation of $\kappa\sub{L}$ in Equation \ref{eqn:kappa_atan}. Notably, the relevant scatterer volume for the primitive unit cell formalism is the volume of the primitive unit cell ($V\sub{uc}$) not the volume per atom ($V_0$), because this model treats one large mass per unit cell\cite{Klemens1955, Tamura1984}. This value enters into the phonon scattering rate (Equation \ref{eqn:tauPD} below) and, therefore, the $\kappa\sub{L}$ prediction. The primitive unit cell formalism for $\Gamma$, while suggested in several seminal papers, is rarely implemented to calculate thermal conductivity\cite{Klemens1958, Klemens1957, Klemens1955, Slack1955, Goldsmid2007, Tamura1984}. Instead, $\Gamma$ is most often computed from Equation \ref{Gamma} based on ``per atom" quantities. In some of these studies, however, the primitive unit cell volume, instead of the volume per atom, is still used to calculate the disorder parameter $u$ in Equation \ref{eqn:kappa_atan}, which leads to an over-prediction of the lattice thermal conductivity reduction (see Supplementary Section \ref{supp:vac_scatt}) proportional to the number of atoms in the unit cell\cite{Meisner1998, Wang2011, Pei2009, Tan2015}. Typically, however, a cancellation of errors prevents the general conclusions of these papers about the importance of point defect scattering from being incorrect\cite{Meisner1998, Tan2015, Pei2009, Wang2011, Qu2011a}. This effect is discussed in greater detail in Section \ref{sct:vacancy} on the large scattering strength of vacancies and interstitial defects.  

\section{Model Derivation: Umklapp and Point Defect Scattering Treatments} 
The following section reviews the full derivation of the Callaway/Klemens model to generalize beyond mass difference scattering alone, clarify assumptions made in the derivation, and discuss the model's dependence on phonon dispersion, a topic garnering recent interest \cite{Hopkins2011, Schrade2018}.

Lattice waves, or phonons, carry a substantial amount of heat through the lattice, characterized by the lattice thermal conductivity ($\kappa\sub{L}$). The efficiency of a phonon with frequency $\omega$ at transporting heat is characterized by its heat capacity ($C\sub{s}$), group velocity ($v\sub{g} = d\omega/dk$), and relaxation time ($\tau$). The lattice thermal conductivity can then be expressed in terms of these values by integrating over phonon frequency up to a maximum frequency supported by the lattice ($\omega\sub{m}$)\cite{Callaway1959, Callaway1960}. If the high temperature approximation is made, the heat capacity at frequency $\omega$ directly relates to the density of states ($g(\omega)$) as $C\sub{s}(\omega) = k\sub{B}g(\omega)$.

\begin{equation}
\kappa\sub{L} = \frac{1}{3}\int_0^{\omega\sub{m}} C\sub{s}(\omega)v\sub{g}(\omega)^2\tau(\omega)d\omega
\label{eqn:kappaL}
\end{equation}

The relaxation time of the phonons is limited by the scattering sources in the crystalline material. Each main source of scattering has an associated relaxation time, including: boundary scattering off of planar defects ($\tau\sub{b}$), umklapp phonon-phonon scattering ($\tau\sub{U}$), and point defect scattering ($\tau\sub{PD}$). Their associated scattering rates are summed according to Matthiessen's rules ($\tau^{-1} = \tau\sub{b}^{-1} + \tau\sub{U}^{-1} + \tau\sub{PD}^{-1}$), assuming that the scattering mechanisms are uncoupled. The model for alloy scattering typically used to describe thermal conductivity versus alloy composition trends neglects boundary scattering to yield the total relaxation time of $\tau = \tau\sub{U}\tau\sub{PD}/(\tau\sub{U} + \tau\sub{PD})$.

The scattering rate from a static imperfection can be derived using Fermi's Golden Rule to define the transition probability ($W_{k,k'}$) from an initial state ($k$) to a final state ($k'$). The transition probability is related to the square of the perturbation matrix element, a measure of the overlap between two phonon states induced by a perturbation to the lattice energy, and includes a lattice energy conservation criteria captured by $\delta(E - E')$. This transition probability is then summed over all the possible final phonon states ($k'$), restricted by the conservation conditions of the phonon interaction.

\begin{equation}
    W_{k,k'} = \frac{2\pi}{\hbar}\bra{\mathbf{k}}H'\ket{\mathbf{k'}}^2 \delta(E - E') 
\end{equation}

The three contributions to point defect scattering (Figure \ref{fig:PDmech}) introduced above are mass contrast ($\Delta M$), force constant contrast ($\Delta K$) and radius contrast ($\Delta R$) and each perturb a different term in the lattice Hamiltonian ($H$)\textemdash the kinetic energy ($T$), harmonic potential energy ($U_\mathrm{2nd}$), and the third order, anharmonic potential energy ($U_\mathrm{3rd}$), respectively \cite{Klemens1955, Carruthers1961, Ortiz2015,Abeles1963,Katre2017, Zhao2017}. The energy perturbation ($H'$) induced by the point defect on site $\mathbf{r}$ with a set of linkages to nearest neighbor sites (${\mathbf{b}_n}$) is then a combination of the effects discussed above

\begin{equation}
    H' = \frac{1}{2}\Delta M \left(\frac{d\textbf{u}(\textbf{r})}{dt}\right)^2 + \sum_n \frac{1}{2}\Delta K_{\textbf{b}_n} [\textbf{u}(\textbf{r}) - \textbf{u}(\textbf{r} - \textbf{b}_n)]^2 + \sum_n\gamma \,    \mathbf{\eta}(\Delta R)[\textbf{u}(\textbf{r}) - \textbf{u}(\textbf{r} - \textbf{b}_n)]^2.
\end{equation}

The scattering due to local, static strain ($\mathbf{\eta}$) depends on the anharmonicity of the distorted bonds, as captured in the Gr\"{u}neisen parameter ($\gamma$). Notably, both the $\Delta K$ and $\Delta R$ effects are captured in the changes to the DFT-calculated interatomic force constants, which change locally in response to both structural relaxation and an altered chemical environment\cite{Katre2017}. Additionally, the strain scattering strength is scaled by the coefficient $Q$, which approximates the number of distorted nearest neighbor bonds around a point defect. Assuming a cubic lattice with a strain field that decays with distance cubed, $Q = 4.2$. If all three effects are present in a system, they combine according to Equation \ref{eqn:fullgamma}, where, as before, atomic species are indexed with $i$ and atomic sites in the unit cell are indexed with $n$ \cite{Klemens1955, Abeles1963}.

\begin{equation}
\Gamma_{n} =  \sum_i f_i \left(\frac{\Delta M^2}{\langle \overline{M} \rangle^2} + 2 \left( \frac{\Delta K}{\langle \overline{K} \rangle} - 2Q\gamma \frac{\Delta R}{\langle \overline{R} \rangle} \right)^2 \right)
\hspace{20 mm}
\Gamma = \langle \Gamma_{n} \rangle
\label{eqn:fullgamma}
\end{equation}


The change in force constant ($\Delta K$) is not an intuitive value, but it is typically assumed that force constants change proportionally with atomic volumes. Therefore, the force constant difference and local strain terms are combined, and both are expressed through the average variance in atomic radius, defined analogously to the mass scattering parameter in Equation \ref{eqn:M2n} and Equation \ref{Gamma}. As before, the atomic radius variance on the $n$\textsuperscript{th} site is defined from the atomic radius of the $i$\textsuperscript{th} species which may occupy that site $R_{i,n}$ and the average atomic radius of the site $\overline{R_n}$. Since the relationships between force constants and atomic volumes are system dependent, these effects are captured in a phenomenological fitting parameter $\epsilon$, which can vary in value on between 1-500 in order to fit to experimental data. 

\begin{equation}
    \Gamma = \frac{\langle \overline{\Delta M^2} \rangle}{\langle \overline{M} \rangle^2} + \epsilon \frac{\langle \overline{\Delta R^2} \rangle}{\langle \overline{R}\rangle^2}
    \hspace{20 mm}
    \langle \Delta R^2 \rangle = \big \langle \sum_i f_i (R_{i,n} - \overline{R_n})^2  \big \rangle 
\end{equation}

For the point defect scattering rate ($\tau\sub{PD}^{-1}$), only two phonon states, an incident and final state, are involved in the interaction. Given the conservation of energy condition, the frequencies of the final and initial phonons are equal. Therefore, the sum over all final scattering states contributes a factor of the 3D density of states ($g$) at the phonon frequency $\omega$\cite{Berman1976}

\begin{align}
\tau\sub{PD}^{-1} = \frac{V\sub{uc} \pi\Gamma \omega^2 g(\omega)}{6},
\hspace{20mm}
g(\omega) = \frac{3\omega^2}{2\pi^2v\sub{p}^2(\omega)v\sub{g}(\omega)}.
\label{eqn:tauPD}
\end{align}

In umklapp scattering, phonons scatter other phonons by virtue of the lattice distortions they generate. The scattering strength is, then, also related to the anharmonicity of the distorted bonds via the Gr\"{u}neisen parameter, $\gamma$, in addition to the phase velocity of the phonon producing the distortion ($v\sub{p}(\omega) = \omega/k$) and the group velocity of the final phonon state ($v\sub{g}(\omega'')$)\cite{Peierls1955,Berman1976}.
Umklapp scattering is referred to as a three-phonon process, and in the typical picture, either two incident phonons combine to form a final phonon state or an initial phonon splits into two final phonon states. Unlike normal three-phonon processes, umklapp processes do not conserve momentum, but instead include an exchange of momentum with the crystal lattice, which given periodicity constraints, must occur in intervals of a reciprocal lattice vector ($b = 2\pi/a$). The relevant conservation law is then: $\mathbf{k} + \mathbf{k'} = \mathbf{k''} + \mathbf{b}$. This more complex conservation law yields a less intuitive set of final available phonon states, and the numerical prefactor of the umklapp scattering rate varies somewhat from source to source, depending on the level of complexity assumed for the expression of final states\cite{Roufosse1976,Toberer2011,Parrott1963,Berman1976}. 

\begin{equation}
\tau\sub{U}^{-1} = \frac{4 \pi a \gamma^2\omega^2k\sub{B}}{\sqrt{2}M v\sub{p}^2(\omega) v\sub{g}(\omega'')}T
\label{eqn:tauU}
\end{equation}

The umklapp scattering rate appears to have the same motif present in the density of states: $\omega^2/(v\sub{p}^2v\sub{g})$. However, due to the increased complexity of the energy and momentum conservation in a 3-phonon process, the sum over final phonon states does not simply contribute a factor of the 1-phonon density of states $g$. Rather, the selection rule for 3-phonon processes is more accurately captured by a calculation of the joint density of states, representing allowed phonon transitions, weighted by the equilibrium occupation numbers of the phonon modes $\mathbf{k'}$ and $\mathbf{k''}$\cite{Chaput2015}. In Klemens, the approximation is made that the magnitude of the final phonon wavevector ($k''$) is small with respect to a reciprocal lattice vector, and as such, the phase space for final phonon states approaches the 1-phonon density of states, contributing a factor of $g(\omega)$ to the umklapp scattering rate\cite{Chaput2015}.

At this point it is typical to make the Debye approximation, which suggests that the $v\sub{g}$ and $v\sub{p}$ are independent of frequency and equal to the classical speed of sound ($v\sub{s} = d\omega/dk|_{k\rightarrow0}$). Equation \ref{eqn:kappaL} for lattice thermal conductivity simplifies, after substituting in the expressions for relaxation time and specific heat, to the integral form of $\mathrm{arctan}$ and gives the final expression for $\kappa\sub{L}$ shown in Equation \ref{eqn:kappa_atan}.

\section{Dispersion Relation Sensitivity}

The formalism above has shown wide applicability to thermoelectric materials, which often have complex, non-Debye dispersions. The examples depicted of Si isotope scattering and \ch{Mg_2Si_{1-x}Sn_x} shown in Figure \ref{fig:fp_comp} both show excellent correspondence between the first principles methods and the Klemens model. Since both materials disagree with the Debye model dispersion implicitly assumed in Klemens theory, the suitability is surprising \cite{Agne2018, Klobes2019}.

Previous studies have explicitly compared the $\kappa\sub{L}$ predictions of the Klemens model using various approximations of the phonon dispersion relation, ranging from the Debye model to polynomial or trigonometric fits of the dispersion\cite{Hopkins2011,Schrade2018}. For example, in a study of two Half-Heusler systems, three different approximations were used to describe the phonon structure of the two materials\textemdash the Debye model, a truncated Debye model, and a cubic polynomial fit of the full dispersion relation. The predicted $\kappa_L$ versus defect concentration curve was plotted for each case and compared to experimental results. The study showed that the prediction of the pure thermal conductivity ($\kappa_0$) depended on the choice of dispersion. However, the ratio $\kappa_L/\kappa_0$ was shown to be independent of the dispersion relation choice, suggesting that while full features of the dispersion relation are required to model the thermal conductivity of pure solids, the suppression of $\kappa\sub{L}$ due to point defects can be described more generally\cite{Schrade2018}.

The dispersion relation dependence enters into the Klemens model through the factors of density of states and the frequency-dependent phonon velocities. In Equation \ref{eqn:kappaL_simp} for lattice thermal conductivity, the relaxation times are re-written to isolate the density of states contribution ($\tau\sub{PD}^{-1} = a\, g(\omega)\omega^4, \tau\sub{U}^{-1} = b\,g(\omega)\,\omega^2$) with coefficients $a$ and $b$ combining any physical and material constants. The factor of $g(\omega)$ cancels in each of the relaxation times as well as the heat capacity, softening the dispersion dependence of the expression. A full derivation of this form is included in Supplementary Section \ref{supp:arctan}.


\begin{align}
     \kappa\sub{L} & = k_B\int_0^{\omega_m} v_g^2(\omega) g(\omega)\omega^2 \frac{(1/b\, g(\omega))\omega^2}{1+a\,g(\omega)\omega^2/b\,g(\omega)} d\omega = \frac{k_B}{b}\int_0^{\omega_m} v_g^2(\omega)  \frac{1}{1+a\omega^2/b} d\omega\,.
     \label{eqn:kappaL_simp}
\end{align}


At this point, the factor of $v_g^2$ remains as a dispersion relation quantity in the model. Therefore, the dispersion dependence is not eliminated from the model, but softened. 
However, the partial cancellation in dispersion relation quantities, particularly stemming from the density of states dependence of several quantities in the model, helps justify this model's application to a wide range of complex, functional electronic materials.

\section{Scattering due to Vacancies and Interstitials}
\label{sct:vacancy}
The Klemens/Callaway model is best defined for randomly dispersed substitutional defects. However, initial work on other off-stoichiometric defects, including vacancies and interstitials, have shown large phonon scattering effects and warrant further investigation.

Vacancy scattering has been demonstrated to induce a large reduction in thermal conductivity in several thermoelectric compounds\cite{Wang2013, Wang2011, Pei2009, Tan2015, Tan2018, Shen2016, Bocher2017, Qu2011a}. In several of these cases, the reduction in $\kappa\sub{L}$ is attributed to mass difference scattering alone. However, it was identified that the volume in Equation \ref{eqn:tauPD} was incorrectly defined as the volume of the unit cell rather than the volume per atom, leading to an over-prediction of the thermal conductivity change\cite{Meisner1998, Tan2015, Pei2009, Wang2011, Qu2011a}.

However, the large perturbation induced by vacancies is still well-described using Klemens theory\cite{Ratsifaritana1987, Klemens1999}. In this case, the lattice energy perturbation comes from missing kinetic energy ($T'$) related to the mass of the removed atom and missing potential energy related to the removed bond between two atoms, or double the potential energy per atom ($2U'$). Within the harmonic approximation ($E=T+U_\mathrm{2nd}$), the kinetic and potential energy perturbations of a single atom should be equal ($T' = U'$) according to the virial theorem, allowing one to relate the potential energy perturbation to the average atomic mass in the lattice ($\left< \overline{M} \right>$). In the calculation of $\Gamma$, the perturbation at a vacancy site can be represented by the mass difference $M_{i,n} - \overline{M_n} = -M\sub{vac} - 2\left< \overline{M} \right>$ in Equation \ref{Gamma} and Equation \ref{eqn:Tamura}, where $M\sub{vac}$ is the mass of the vacant atom\cite{Ratsifaritana1987, Klemens1999}. 

\begin{figure}
    \centering
    \includegraphics[width = \textwidth]{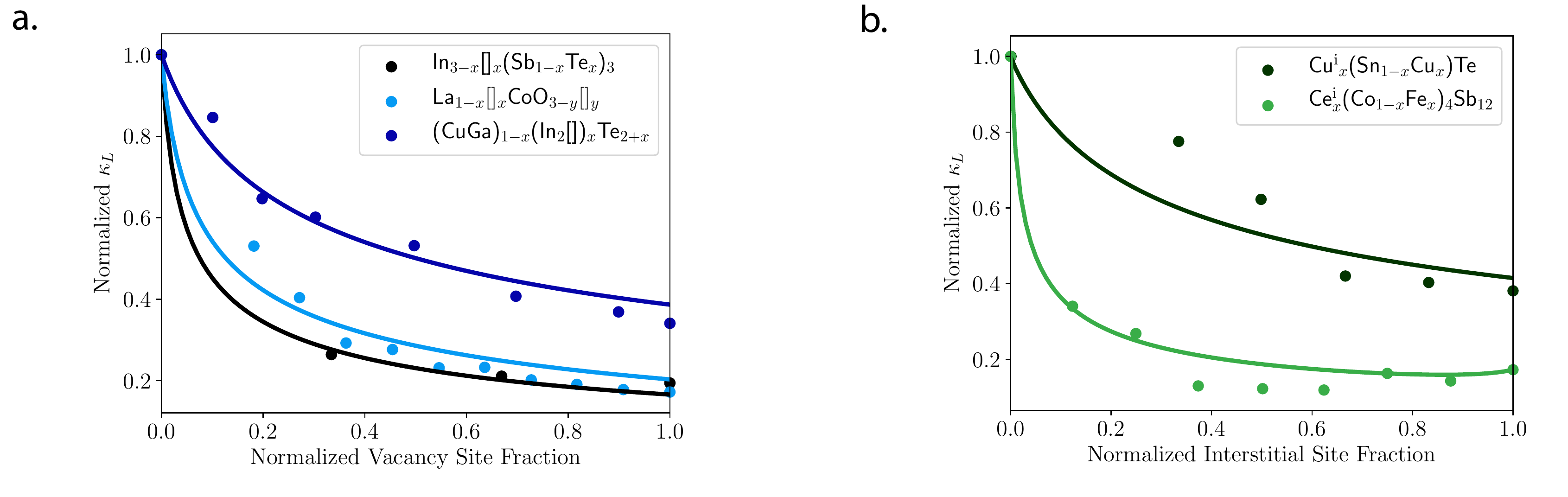}
    \caption{Both vacancy and interstitial scattering data from literature (points) can be described using a simple treatment of broken (or added) bonds based on the virial theorem (line). Normalized thermal conductivity reductions for systems with (a) stoichiometric vacancies, where [] represents a vacancy \cite{Wang2011, Shen2016, Pei2009} and (b) stoichiometric interstitial atoms \cite{Meisner1998, Pei2016}}
    \label{fig:vac_int_scatt}
\end{figure}

This simple treatment of vacancy scattering performs well in many defective solids, some of which are reproduced in Figure \ref{fig:vac_int_scatt}a. The experimental data shown would not be described by standard mass difference alone and requires the perturbation induced by a missing bond. The Supplementary Section \ref{supp:vac_scatt} compares results for the mass difference only curve versus the full inclusion of the broken bonds term, and depicts how an incorrect definition of volume can lead to a cancellation of errors. 


The suitability of the vacancy model suggests, then, that interstitial atoms may be describable with an identical treatment. Interstitial or filler atoms represent the reverse situation, where an extra mass ($M\sub{int}$) is added onto a site and a new bond forms between the interstitial atom and a neighbor; therefore, a perturbation of $T' + 2U'$ should apply, yielding essentially the same mass difference as before ($M_{i,n} - \overline{M_n} = M\sub{int} + 2\left< \overline{M} \right>$). It should be noted that the interstitial atom sites have a stoichiometry corresponding to the ratio of interstitial to lattice sites. While interstitial scattering requires more detailed study across additional materials systems, the initial data represented in Figure \ref{fig:vac_int_scatt}b, supports the application of the virial theorem treatment. 



\FloatBarrier

\FloatBarrier

\section{Conclusion}

The analytical point defect scattering model provides a simple route to identify scattering mechanisms in a system. In several systems, comparisons of alloy scattering models with different scattering terms excluded, such as strain scattering or broken bonds, provide a lens into the most potent scattering effects and a route to optimally tailor the thermal properties via defect engineering\cite{Pei2016, Zheng2018, Tan2015, Meisner1998}. The systems best described by this model are those with well-defined crystal structures, randomly dispersed point defects, and low magnitude perturbations, such that regions of high mass contrast and high defect concentration may require verification\cite{Shiga2014}. However, the large thermal conductivity reduction induced by vacancies and interstitial atoms is still described by these analytical equations using the virial theorem to model the perturbation due to the formation or removal of nearest neighbor bonds. 


In addition, the suitability of the alloy model for arbitrary dispersion relations suggests that the ratio of alloy scattering to umklapp scattering predicted by the model is fairly dispersion relation independent. This reduced sensitivity to dispersion can be understood through a partial cancellation of the density of states in the phonon relaxation times and heat capacity. As a result, it is found that these equations are justifiable in describing the impact of point defects on the thermal properties of materials with complex atomic and phonon structures attracting attention in fields like thermoelectrics and microelectronics.

\section*{Acknowledgements}
We thank NSF DMREF award \#1729487 and award \#70NANB19H005 from U.S. Department of Commerce, National Institute of Standards and Technology as part of the Center for Hierarchical Materials Design (CHiMaD) for support.


\bibliography{PDPaper}

\vspace{20 mm}
\section{Supplementary Information}

\subsection{Tamura Model Low Frequency Limit}
\label{supp:lf}
This section describes the low frequency approximation of the Tamura model. The Tamura expression with full structural and lattice dynamical dependence is shown in Equation \ref{eqn:Tamura} and reproduced below.

\begin{equation}
\Gamma^\mathrm{T}\sub{M} = \sum_s \sum_i f_{i,s}(\frac{M_{i,s} - \overline{M_s}}{\overline{M_s}})^2|(\mathbf{e_k}(s)\cdot \mathbf{e_{k'}}(s))|^2
\label{eqn:suppTamura}
\end{equation}

In addition, an approximation is described in the low frequency limit to yield a closed form expression, which depends only on the atomic masses. The assumption made here is that the displacement ($\mathbf{u}$) of each atom in a low-frequency mode is roughly equal in magnitude; therefore, the magnitude of an eigenvector element is proportional to the square root of the atomic mass ($|\mathbf{e}(\mathbf{k}, s)| \propto \sqrt{M_s}$). This suggests that the squared polarization vector dot product ($|(\mathbf{e_k}(s)\cdot \mathbf{e_{k'}}(s))|^2$) weights the mass difference on a site depending on its mass relative to the other atoms in the formula unit, or an approximate factor of ($\overline{M_s}^2/\langle \overline{M} \rangle ^2$). This treatment results in Equation \ref{eqn:M2n} and \ref{Gamma} suggested in the Introduction, as depicted below.

\begin{equation}
    \Gamma^\mathrm{lf}\sub{M} = \frac{1}{\langle \overline{M} \rangle^2}\frac{\sum_n c_n (\overline{M_n})^2 \sum_i f_i (1 - M_{i,n}/\overline{M_n})^2}{\sum_n c_n}= \frac{\langle \overline{\Delta M^2} \rangle}{\langle \overline{M}\rangle^2}  
\end{equation}

In the original paper by Tamura, the low frequency limit instead as:

\begin{equation}
    \Gamma^\mathrm{lf}\sub{M} = \frac{\sum_n c_n (\overline{M_n})^2 \sum_i f_i (1 - M_{i,n}/\overline{M_n})^2}{\sum_n c_n \overline{M_n}^2} = \left(\frac{1}{\langle \overline{M}^2 \rangle}\right)\frac{\sum_n c_n \Delta M^2_n}{\sum_n c_n}
    \hspace{20mm}
    \langle \overline{M}^2 \rangle= \frac{\sum_n c_n \overline{M_n}^2}{\sum_n c_n}.
    \label{eqn:suppTamuraLF}
\end{equation}

\begin{figure}
    \centering
    \includegraphics[width=0.6\textwidth]{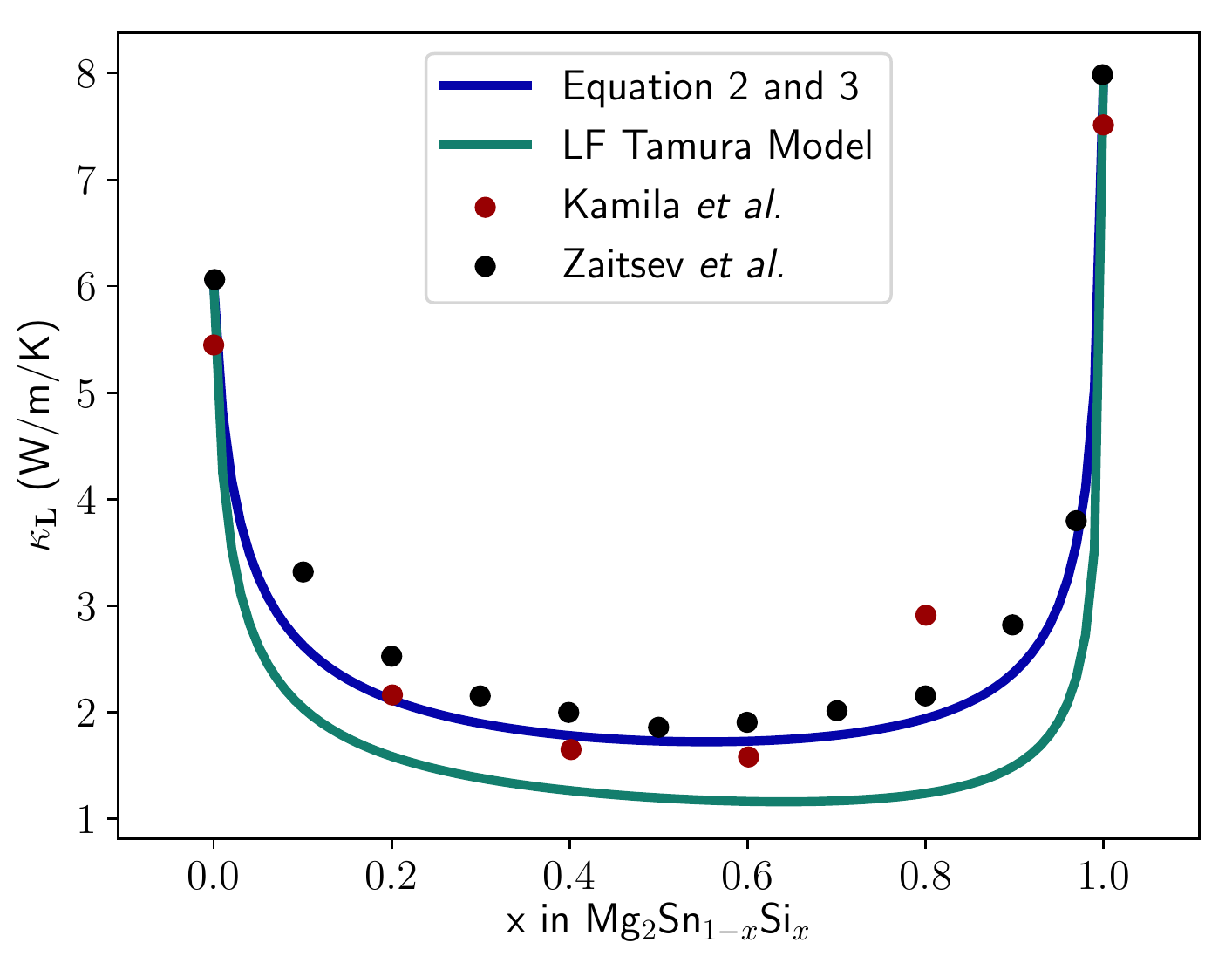}
    \caption{The predictions of the low frequency Tamura model and Equation \ref{eqn:M2n} and \ref{Gamma} in the paper are compared for the example of point defect scattering in the  \ch{Mg_2Sn} and \ch{Mg_2Si} solid solution. The low frequency Tamura model typically over-predicts the thermal conductivity reduction due to point defect scattering.}
    \label{supp:tam_comp}
\end{figure}

As shown, this expression can be rearranged to a form similar to Equation \ref{Gamma}. However, the expression is subtly different, as it includes an averaging of the squared atomic masses ($\langle\overline{M_n}^2\rangle$), rather than the average mass, which is finally squared ($\langle \overline{M_n}\rangle^2$). Through comparison with experimental and simulated thermal conductivity data, the mass difference term provided in the Introduction (Equation \ref{Gamma}) is verified to give more accurate predictions, while Equation \ref{eqn:suppTamuraLF} can deviate by 30-40\%. Figure \ref{supp:tam_comp} compares the results of the low frequency Tamura model and Equation \ref{Gamma} from in Introduction for the \ch{CaZn_2Sb_2} and \ch{YbZn_2Sb_2} solid solution. Moreover, the mass difference parameter from Equation \ref{eqn:suppTamuraLF} suffers from a lack of generalizability to arbitrary unit cell sizes, such that a doubling of the unit cell leads to an increase in the predicted scattering rate.

\subsection{Derivation of Relaxation Times}
\label{supp:tau}
This section details the derivations of the point defect and umklapp relaxation times from Fermi's Golden Rule to understand their dependence on phonon velocity and dispersion relation.

\subsubsection{Point Defect Scattering}
Point defects act as a static perturbation, and therefore the scattering rate can be determined using Fermi's Golden Rule based on first order perturbation theory. Previous work has shown that higher order perturbation terms have a negligible effects on the lattice energy\cite{Tamura1983, Larkin2017}. The probability of scattering from state $\mathbf{k}$ to $\mathbf{k'}$ (W$_{\mathbf{k,k'}}$) is proportional to the square of the perturbation matrix element, a measure of the overlap between two phonon states due to a perturbation to the lattice energy, with a conservation of energy enforced through the $\delta(\omega_{k} - \omega_{k'})$ term.

\begin{align}
    &W_{k,k'} = \frac{2\pi}{\hbar^2}\bra{\mathbf{k}}H'\ket{\mathbf{k'}}^2 \delta(\omega_k - \omega_{k'}) 
    \label{eqn:w}
\end{align}


The perturbation matrix element includes a coefficient ($C$), which captures the physics of the perturbation induced by the point defect, while $a(k)$ and $a^{*}(k')$ are creation and annihilation operators to represent the change in occupation numbers of the $\mathbf{k}$ and $\mathbf{k'}$ states as a result of the phonon-impurity interaction.
\begin{equation}
    \bra{k}H'\ket{k'} = C(k,k')a(k)a^{*}(k')
\end{equation}

Substituting in the full form of the creation and annihilation operators gives the expression below, where $N$ refers to the number of phonons in a given state. 

\begin{equation}
    \bra{k}H'\ket{k'}^2 = \frac{\hbar^2}{M^2\omega^2}C^2(k,k')[N(N'+1) - N'(N+1)]
\end{equation}

It has been shown that the term in the square brackets reduces to 1 in the integral over the constant energy surface corresponding to final $\mathbf{k'}$ states\cite{Klemens1955}. 

Here, the coefficient ($C$) will be calculated for the mass difference case, in which the perturbation stems from an atom with a mass of $M' = M_0 + \Delta M$ sitting in the primitive unit cell located at $\mathbf{R}$. The perturbation due to force constant fluctuation and strain are similar in form. The energy perturbation ($E'$) to the lattice due to mass difference comes in through the kinetic energy term, where $\dot{u}(\mathbf{R})$ signifies the time derivative of the unit cell displacement. 

\begin{equation}
E'(\mathbf{R}) = \frac{1}{2}\Delta M(\mathbf{R}) \dot{u}^2(\mathbf{R})
\end{equation}

The real space perturbation is written in terms of a reciprocal space vector ($\mathbf{Q}$) by taking the Fourier transform. Here, $S$ refers to the number of primitive unit cells in the lattice\cite{Tamura1983, Tamura1984}. 

\begin{equation}
 \Delta \Tilde{M}(\mathbf{Q}) = \frac{1}{S} \sum_{\mathbf{R}} \Delta M(\mathbf{R}) e^{i\mathbf{QR}}
\end{equation}

The expression for $C^2$ picks up a factor of $\Delta \Tilde{M}(\mathbf{Q})\Delta \Tilde{M}(\mathbf{Q'})$, which is equal to:

\begin{equation}
\Delta \Tilde{M}(\mathbf{Q})\Delta \Tilde{M}(\mathbf{Q'}) = \frac{1}{S^2} \sum_\mathbf{R, R'} \Delta M(\mathbf{R})\Delta M(\mathbf{R'})e^{i(\mathbf{Q'R' - QR})}    
\end{equation}

If the approximation is made that the point defects are randomly distributed over the lattice, the sum over lattice sites can instead be written as an average squared mass difference ($\Delta M^2$) times the number of defect sites in the lattice ($S_i$)\cite{Tamura1983,Tamura1984}.

\begin{align}
\Delta \Tilde{M}(\mathbf{Q})\Delta \Tilde{M}(\mathbf{Q'}) &= \frac{1}{S}\frac{S_i}{S}\Delta M ^2 = \frac{1}{S}f_i\Delta M ^2 
\end{align}

The velocity $\dot{u}$ is again written in terms of creation and annihilation operators, which contribute a frequency and polarization vector dependence to yield the full expression for $C^2$ \cite{Klemens1958}.

\begin{equation}
    C^2 = \frac{1}{4S}f_i(\Delta M)^2\omega^4|(\mathbf{e_k}(s)\cdot \mathbf{e_{k'}}(s))|^2
    \label{eqn:c2}
\end{equation}

Equation \ref{eqn:c2} can then be substituted into Equation \ref{eqn:w} for the transition probability to get the simplified expression shown below.

\begin{equation}
   W_{k,k'} =\frac{\pi}{2S}f_i\frac{\Delta M^2}{M^2}\omega^2\delta(\Delta \omega)|(\mathbf{e_k}(s)\cdot \mathbf{e_{k'}}(s))|^2
   \label{eqn:w_simp}
\end{equation}

To get the scattering rate, one must then sum $W_{k,k'}$ over all the possible final phonon states $\mathbf{k'}$. Given the conservation of momentum constraint ($\left|\mathbf{k}\right| = 
\left|\mathbf{k}'\right|$), this constitutes an integral over a constant energy surface or sphere of radius $k$ in k-space. In the conversion from a sum over discrete $k$ states to an integral over $k$ states, a volume factor of $V\sub{tot}/(2\pi)^3$ is picked up, where $V\sub{tot}$ is the volume of the crystal.

\begin{equation}
    \tau\sub{PD}^{-1} = \frac{V\sub{tot}}{(2\pi)^3} \int W_{k,k'} d^3\mathbf{k'}
\end{equation}

The spherical surface integral is evaluated noting that: (1) $d^3\mathbf{k'} = k'^2\mathrm{sin}\Theta dk d\Theta d\phi$, (2) $\omega_k = \omega_{k'} = v_p(\omega)k'$, (3) $\int {sin}\Theta dk d\Theta d\phi = 4\pi$, (4) $\delta(\Delta \omega) = \delta(\Delta k)/v_g(\omega)$ and (5) $V\sub{uc} = V\sub{tot}/S$ is the volume of the primitive unit cell.

\begin{equation}
    \tau\sub{PD}^{-1} = \frac{V\sub{uc}}{4\pi}f_i \left(\frac{\Delta M}{M} \right)^2|(\mathbf{e_k}(s)\cdot \mathbf{e_{k'}}(s))|^2\frac{\omega^4}{v_p^2(\omega)v_g(\omega)}
\end{equation}

Finally, the relaxation time can be written in terms of the 3D phonon density of states ($g(\omega)$) given in Equation \ref{eqn:tauPD}.  
\begin{equation}
    \tau\sub{PD}^{-1} = \frac{\pi V\sub{uc}}{6} f_i \left(\frac{\Delta M}{M} \right)^2|(\mathbf{e_k}(s)\cdot \mathbf{e_{k'}}(s))|^2g(\omega)\omega^2   
\end{equation}

As discussed before in relation to the Tamura model (Section \ref{sct:Tamura}), the scattering parameter $\Gamma = f_i \left(\Delta M/M \right)^2|(\mathbf{e_k}(s)\cdot \mathbf{e_{k'}}(s))|^2 $, which reproduces Equation \ref{eqn:tauPD} in the main text.

\begin{equation}
\tau\sub{PD}^{-1} = \frac{V\sub{uc} \pi\Gamma \omega^2 g(\omega)}{6}
\end{equation}

\subsubsection{Umklapp Scattering}

The umklapp scattering rate follows a similar derivation, however, the conservation rule is more complicated since the process involves three phonon modes ($\mathbf{k,k',k''}$) as well as momentum exchange with the lattice via the addition or subtraction of a reciprocal lattice vector ($\mathbf{b}$). The derivation here is adopted from the derivation by Klemens for strain scattering off a point defect, however in this case, the strain is produced by another phonon rather than a point imperfection \cite{Klemens1958}.

\begin{equation}
    \mathbf{k} + \mathbf{b} = \mathbf{k'} + \mathbf{k''}
    \label{eqn:Uconserve}
\end{equation}

The umklapp perturbation matrix element is similar in form to that of point defect scattering, but now includes three creation or annihilation operators since the process involves the change in occupation number for three phonon modes. Additionally, the coefficient ($C\sub{U}$) is dependent on the anharmonicity of the lattice. 

\begin{equation}
    \bra{i}H'\ket{f}^2 = [C\sub{U}(k,k',k'')a(k)a^{*}(k')a^{*}(k'')]^2 = \frac{\hbar^3}{M^3\omega\omega'\omega''}C\sub{U}^2(k,k', k'')[(N+1)(N'+1)(N'' +1) - NN'N'']
    \label{eqn:umklapp}
\end{equation}

In Klemens, the phonon mode $k''$ is treated as a Fourier strain component producing a perturbation to the lattice energy. If a uniform dilatational strain ($\Delta$) is assumed, the Fourier component can be written as $i \omega''/v_p(\omega'')\sqrt{S}$ in the limit $k'' \rightarrow 0$. The elastic strain impacts force constants, and therefore induces a frequency shift modeled using the Gr\"uneisen model \cite{Klemens1958}.

\begin{equation}
    \omega{\mathbf{k}} = \omega_0(\mathbf{k})[1 - \gamma(\mathbf{k})\Delta]
    \label{eqn:grunmodel}
\end{equation}

The coefficient $C\sub{U}$ then represents the lattice energy change associated with a uniform dilatational strain, defined using the Gr\"uneisen model, and the form of the uniform dilatation as $k''$ approaches 0.

\begin{equation}
    C\sub{U}(k, k', k'') = \frac{-2i}{\sqrt{S}v_p(k'')}\gamma M \omega \omega' \omega''
    \label{eqn:cu_full}
\end{equation}

The final component of the squared matrix element (shown in Equation \ref{eqn:umklapp}) is the term in the square brackets, representing the difference in occupation of phonon modes from the initial to final state. At the high temperature limit, this term can be written in terms of the Bose-Einstein distribution such that it reduces to: $k_BT\omega/\hbar \omega'\omega''$. The full form of the squared matrix element simplifies to the form shown below. 

\begin{equation}
   \bra{i}H'\ket{f}^2 = \frac{\hbar^2}{M}\frac{4\gamma^2\omega^2}{Sv_p^2(k'')}k_BT
    \label{eqn:sqHprime_full}
\end{equation}

Just as before, the scattering probability is defined using Fermi's Golden Rule (Equation \ref{eqn:w}, where the initial and final states are now represented as $\ket{i}$ and $\ket{f}$, for simplicity. As before, the scattering rate is calculated by summing over $W_{i,f}$ for all possible final states. This is achieved by performing a sum over all $\mathbf{k'}$ and $\mathbf{b}$, which then fixes the value of $\mathbf{k''}$ as a result of the conservation condition (Equation \ref{eqn:Uconserve}). 

\begin{equation}
 \tau\sub{U} = \sum_{\mathbf{k'}, \mathbf{b}} W_{i,f}
\end{equation}

It is assumed that $\mathbf{k'}$ is restricted to spheres of radius $\frac{1}{2}(\mathbf{k} + \mathbf{b}$), which is suggested to be true as long as the dispersion relation is not modified by the zone structure\cite{Roufosse1976}. Therefore, the sum can be once again replaced by a surface integral over this sphere, and picks up a volume factor of $V\sub{tot}/(2\pi^3)$, where $V\sub{tot}$ is the volume of the crystal. 

\begin{equation}
    \tau\sub{U}^{-1} = \sum_{\mathbf{b}} \frac{V\sub{tot}}{(2\pi)^3} \int W_{i,f} d^3\mathbf{k'}
    \label{eqn:tauInt}
\end{equation}

Following the same integral simplifications discussed in the derivation of $\tau\sub{PD}$, the scattering rate due to umklapp processes is shown below.

\begin{equation}
    \tau\sub{U}^{-1} = \frac{V\sub{uc}{\pi}\gamma^2\omega^2}{M v_p^2(\omega'') v_g(\omega')}\sum_{\mathbf{b}} (k+b)^2
    \label{eqn:tauEval}
\end{equation}

Finally, the approximation is made that $k$ is small in magnitude in comparison to the reciprocal lattice vector $b = 2\pi/a$, such that $(k+b)^2 = 4\pi^2/a^2$. For a cubic close-packed material with a rhombohedral primitive unit cell, the volume $V\sub{uc}$ is $a^3(\sqrt{2})^{-1}$, and the scattering rate reduces to the form shown below.

\begin{equation}
    \tau\sub{U}^{-1} = \frac{4 \pi a \gamma^2Tk_B}{\sqrt{2}M}\frac{\omega^2}{ v_p^2(\omega) v_g(\omega'')}
\end{equation}

\subsection{Derivation of the Arctan Equation}
\label{supp:arctan}
The lattice thermal conductivity with umklapp and point defect scattering simplifies to a popular function with the form $\mathrm{tan}^{-1}$. The derivation of this form is shown here to highlight the cancellation of phonon velocities in the relaxation times.

Equation \ref{eqn:kappaL} of the text gives an expression for the lattice thermal conductivity in terms of the frequency-dependent heat capacity, phonon group velocity, and phonon lifetime, which is reproduced below. 

\begin{equation}
\kappa = \frac{1}{3}\int_0^{\omega\sub{m}} C\sub{s}(\omega)v\sub{g}(\omega)^2\tau(\omega)d\omega
\label{eqn:kappaL_supp}
\end{equation}

for lattice thermal conductivity can be simplified at the high temperature limit to the following form. 

\begin{equation}
    \kappa = \frac{k_B}{2 \pi^2}\int_0^{\omega\sub{max}} \frac{\omega^2}{v_p^2(\omega) v_g(\omega)}v_g^2(\omega)\frac{\tau_U}{1 + \tau_U/\tau_{PD}} d\omega
\end{equation}

Next, the phonon velocities are pulled out of the coefficients of the relaxation times ($A = a(v_p^2v_g)^{-1}, B = b(v_p^2v_g)^{-1}$).

\begin{equation}
     \kappa = \frac{k_B}{2\pi^2}\int_0^{\omega\sub{max}} v_g^2(\omega) \frac{1}{v_g(\omega) v_p^2(\omega)}\omega^2 \frac{(1/b(v_p^2v_g)^{-1})\omega^2}{1+a(v_p^2v_g)^{-1}\omega^2/b(v_p^2v_g)^{-1}} d\omega  
\end{equation}

The phonon velocities in the specific heat, umklapp relaxation time, and the point defect relaxation time will cancel to yield the final simplified form.

\begin{equation}
      \kappa = \frac{k_B}{2\pi^2b} \int_0^{\omega\sub{max}} v^2_g (\omega) \frac{1}{1+a\omega^2/b} d\omega 
      \label{eqn:kappaComb}
\end{equation}

If the approximation can be made that the factor of $v_g^2$ is largely frequency-independent, then the integral above has a solution in the form of $\mathrm{tan}^{-1}$. 

\begin{equation}
    \kappa = \frac{k_B v_g^2}{2\pi^2b(b/a)^{1/2}}\mathrm{tan}^{-1}(\frac{\omega\sub{max}}{(b/a)^{1/2}})
\end{equation}

The pure lattice thermal conductivity $\kappa_0$ can be calculated from Equation \ref{eqn:kappaComb} when the point defect scattering coefficient $a$ is set to 0. The resulting value of the pure lattice thermal conductivity is: $\kappa_0 = (k_Bv_g^2\omega\sub{max})/(2\pi^2b)$. Therefore, the ratio of $\kappa\sub{D}/\kappa_0$ simplifies to the following form:

\begin{equation}
   \frac{\kappa}{\kappa_0} =  \frac{\mathrm{tan}^{-1}u}{u}
   \hspace{20mm}
   u = \frac{\omega\sub{max}}{(b/a)^{1/2}}
\end{equation}

\subsection{Primitive Unit Cell Mass Model}
\label{supp:puc}
The primitive unit cell mass difference model describes a system in which all individual atom sites in a primitive unit cell are coarse-grained into a single, vibrating mass, and is signified in the main text as $\Gamma\sub{PUC}$. Therefore, all quantities are defined on a per-unit-cell basis. The parameter can be defined through a statistical mechanics model, where the lattice can be described as a sum of microstates, which represent the unit cells in the lattice. Therefore, the full mass difference parameter is determined by taking the mass difference of each microstate weighted by the probability of finding that microstate in the lattice ($P_c$). 

\begin{equation}
\Gamma = \sum_{\mathrm{mic}}P\sub{mic}\Gamma\sub{mic}
\end{equation}

In the case that impurities randomly distribute on a given sublattice, the probabilities can be calculated using the binomial distribution theorem. As an example, say there is a host compound \ch{A_xB_yC_z} with the impurities A', B', and C', which substitute on the A, B, and C sublattice, respectively. If $f_i$ is the atomic site fraction of the i\textsuperscript{th} impurity as before, then the overall composition of the alloy is given by: $\mathrm{[A(1- f_a)A'(f_a)]_x[B(1-f_b)B'(f_b)]_y[C(1-f_c)C'(f_c)]_z}$.

The various microstates that may compose this lattice can be defined by all the possible fillings of the A, B, and C sublattice with host atoms versus impurity atoms. Thus, each sublattice is treated as a binomial distribution in which a number of sites (set by the stoichiometry) can each have one of two outcomes: the site can be occupied by an impurity atom with a probability of $f_i$ or it can be occupied by a host atom with a probability of (1-$f_i$). By the binomial distribution formula, the probability that $r$ of the total $x$ sites on the A sublattice will be replaced by impurity atoms is given by: $\binom{x}{r}f_A^{r}(1-f_A)^{x-r}$.

Now, it is possible to consider the probability of an example microstate ($P
\sub{mic}$), such as a unit cell with 2 impurity atoms on the A sublattice, 1 impurity atom on the B sublattice, and no impurity atoms on the C sublattice. This full probability would have the form below, where the shorthand $P_A(2)$ refers to the probability of having 2 impurity atoms on the A sublattice. 

\begin{align}
    &P\sub{mic} = P_A(2)*P_B(1)*P_C(0)\\
    &P\sub{mic} = [\binom{x}{2}f_A^{2}(1-f_A)^{x-2}]*[\binom{y}{1}f_B^{1}(1-f_B)^{y-1}]*[\binom{z}{0}f_C^{0}(1-f_C)^{z}]
\end{align}

The $\Gamma$ associated with that microstate would be based on the difference between the mass of that specific microstate ($M_{m}$) and the average mass of a unit cell in the lattice. So, for the example microstate above, $\Gamma$ would have the following form:

\begin{align}
    &\Gamma\sub{mic} = \left(1-\frac{M\sub{mic}}{\overline{M}} \right)^2\\
    &M\sub{mic} = M_A(x-2) + M_{A'}(2) + M_B(y-1) + M_{B'} + M_C(z)
\end{align}

\subsection{Elaborated Example of Vacancy Scattering}
\label{supp:vac_scatt}
This section provides a full example of the vacancy scattering model applied to literature values. The thermal conductivity measurements from Wang \textit{et al.} are utilized for \ch{La_{1-x}CoO_{3-y}} with La and O vacancies. The mass difference scattering strength is given by the expression below, where $\overline{M_1}$ is the average mass of the La site and $\overline{M_3}$ is the average mass of the O site. Here, the average atomic mass in the compound $\langle \overline{M} \rangle = (\overline{M_1} + M\sub{Co} + 3\overline{M_3})/5$.

\begin{equation}
\Gamma\sub{M} = \frac{(1/5)(x(0-\overline{M_1})^2 + (1-x)(M\sub{La} - \overline{M_1})^2 + 3(y(0-\overline{M_3})^2 + (1-y)(M\sub{O} - \overline{M_3})^2))}{\langle \overline{M} \rangle^2}    
\end{equation}

In the original text, the full thermal conductivity reduction is explained using mass difference scattering alone, without the perturbation due to broken bonds. However, in this case, the volume in Equation \ref{eqn:tauPD} was incorrectly defined as the volume of the primitive unit cell when defining the point defect relaxation, where it should have been defined as the volume per atom. This error compensates for the missing broken bonds term, such that the curve reported in the paper still adequately represents the data, and the main conclusions about the point defect scattering strength hold. However, using the virial theorem treatment, the above equation can be adjusted by tripling the mass difference on both vacancy sites as shown below.

\begin{equation}
\Gamma\sub{M} = \frac{(1/5)(x(-M\sub{La}-2\langle \overline{M} \rangle)^2 + (1-x)(M\sub{La} - \overline{M_1})^2 + 3(y(-M\sub{O}-2\langle \overline{M} \rangle)^2 + (1-y)(M\sub{O} - \overline{M_3})^2))}{\langle \overline{M} \rangle^2}    
\end{equation}

\begin{figure}[t]
    \centering
    \includegraphics[width=0.6\textwidth]{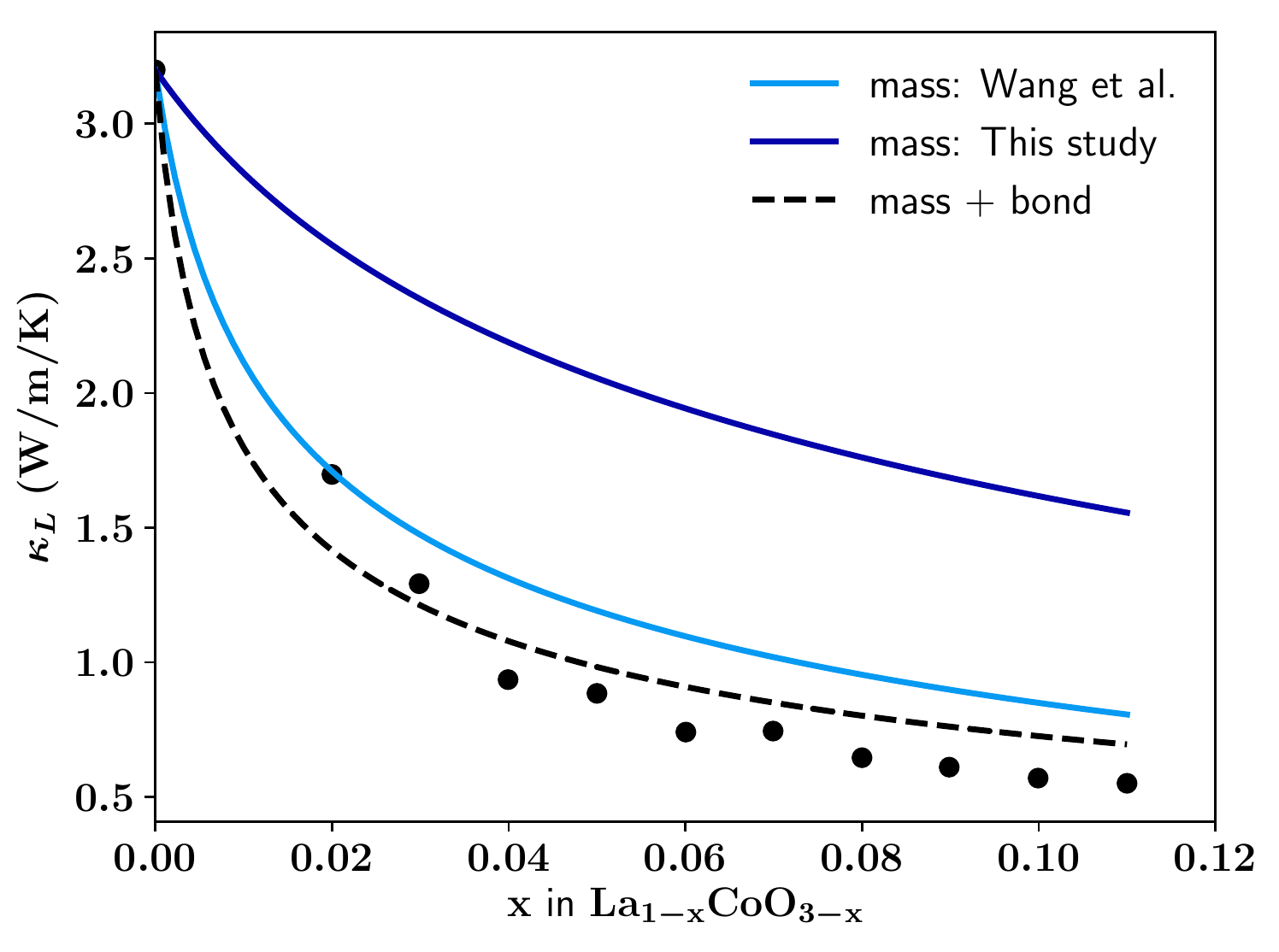}
    \caption{The mass difference model with the unit cell volume error is compared to the mass difference only model described in this study as well as the vacancy model with the virial treatment for broken bonds, which provides an excellent fit\cite{Wang2011}.}
    \label{fig:supp_vac}
\end{figure}

Figure \ref{fig:supp_vac} includes: 1) the model in the original paper using the unit cell volume ($V\sub{uc}$), 2) the revised mass difference only model where the volume per atom ($V_0$) is used, and 3) the model with the virial theorem treatment for broken bonds, where $V_0$ is used. As in the original paper by Wang \textit{et al.}, it is assumed that $x=y$ in the defected chemical formula\cite{Wang2011}.

\end{document}